\documentclass[aps,reprint,superscriptaddress,nofootinbib,amsmath,amssymb,prd]{revtex4-2}

\usepackage{graphicx}
\usepackage{booktabs}
\usepackage[hidelinks,colorlinks=true,linktoc=page,citecolor=blue,filecolor=blue,urlcolor=blue,linkcolor=blue]{hyperref} 
\usepackage{multirow}
\usepackage[normalem]{ulem}
\usepackage{xcolor}
\usepackage{adjustbox}

\usepackage{orcidlink}

\usepackage[nameinlink,noabbrev]{cleveref}

\crefname{equation}{Eq.}{Eqs.}
\crefname{section}{Section}{Sections}
\crefname{figure}{Figure}{Figures}
\crefname{table}{Table}{Tables}
\crefname{appendix}{Appendix}{Appendices}
\Crefname{figure}{Figure}{Figures}
\Crefname{equation}{Equation}{Equations}
\Crefname{section}{Section}{Sections}
\Crefname{table}{Table}{Tables}

\begin{document} 

\title{Early time solution as an alternative to the late time evolving dark energy with DESI DR2 BAO}
\preprint{xxx}

\author{E.~Chaussidon\orcidlink{0000-0001-8996-4874}}
\email{echaussidon@lbl.gov}
\affiliation{Lawrence Berkeley National Laboratory, 1 Cyclotron Road, Berkeley, CA 94720, USA}

\author{M.~White\orcidlink{0000-0001-9912-5070}}
\affiliation{Department of Physics, University of California, Berkeley, 366 LeConte Hall MC 7300, Berkeley, CA 94720-7300, USA}
\affiliation{University of California, Berkeley, 110 Sproul Hall \#5800 Berkeley, CA 94720, USA}

\author{A.~de~Mattia\orcidlink{0000-0003-0920-2947}}
\affiliation{IRFU, CEA, Universit\'{e} Paris-Saclay, F-91191 Gif-sur-Yvette, France}

\author{R.~Gsponer\orcidlink{0000-0002-7540-7601}}
\affiliation{Institute of Physics, Laboratory of Astrophysics, \'{E}cole Polytechnique F\'{e}d\'{e}rale de Lausanne (EPFL), Observatoire de Sauverny, Chemin Pegasi 51, CH-1290 Versoix, Switzerland}

\author{S.~Ahlen\orcidlink{0000-0001-6098-7247}}
\affiliation{Physics Dept., Boston University, 590 Commonwealth Avenue, Boston, MA 02215, USA}

\author{D.~Bianchi\orcidlink{0000-0001-9712-0006}}
\affiliation{Dipartimento di Fisica ``Aldo Pontremoli'', Universit\`a degli Studi di Milano, Via Celoria 16, I-20133 Milano, Italy}
\affiliation{INAF-Osservatorio Astronomico di Brera, Via Brera 28, 20122 Milano, Italy}

\author{D.~Brooks}
\affiliation{Department of Physics \& Astronomy, University College London, Gower Street, London, WC1E 6BT, UK}

\author{T.~Claybaugh}
\affiliation{Lawrence Berkeley National Laboratory, 1 Cyclotron Road, Berkeley, CA 94720, USA}

\author{S.~Cole\orcidlink{0000-0002-5954-7903}}
\affiliation{Institute for Computational Cosmology, Department of Physics, Durham University, South Road, Durham DH1 3LE, UK}

\author{A.~Cuceu\orcidlink{0000-0002-2169-0595}}
\affiliation{Lawrence Berkeley National Laboratory, 1 Cyclotron Road, Berkeley, CA 94720, USA}

\author{A.~de la Macorra\orcidlink{0000-0002-1769-1640}}
\affiliation{Instituto de F\'{\i}sica, Universidad Nacional Aut\'{o}noma de M\'{e}xico,  Circuito de la Investigaci\'{o}n Cient\'{\i}fica, Ciudad Universitaria, Cd. de M\'{e}xico  C.~P.~04510,  M\'{e}xico}

\author{P.~Doel}
\affiliation{Department of Physics \& Astronomy, University College London, Gower Street, London, WC1E 6BT, UK}

\author{S.~Ferraro\orcidlink{0000-0003-4992-7854}}
\affiliation{Lawrence Berkeley National Laboratory, 1 Cyclotron Road, Berkeley, CA 94720, USA}
\affiliation{University of California, Berkeley, 110 Sproul Hall \#5800 Berkeley, CA 94720, USA}

\author{A.~Font-Ribera\orcidlink{0000-0002-3033-7312}}
\affiliation{Institut de F\'{i}sica d’Altes Energies (IFAE), The Barcelona Institute of Science and Technology, Edifici Cn, Campus UAB, 08193, Bellaterra (Barcelona), Spain}

\author{J.~E.~Forero-Romero\orcidlink{0000-0002-2890-3725}}
\affiliation{Departamento de F\'isica, Universidad de los Andes, Cra. 1 No. 18A-10, Edificio Ip, CP 111711, Bogot\'a, Colombia}
\affiliation{Observatorio Astron\'omico, Universidad de los Andes, Cra. 1 No. 18A-10, Edificio H, CP 111711 Bogot\'a, Colombia}

\author{E.~Gaztañaga}
\affiliation{Institut d'Estudis Espacials de Catalunya (IEEC), c/ Esteve Terradas 1, Edifici RDIT, Campus PMT-UPC, 08860 Castelldefels, Spain}
\affiliation{Institute of Cosmology and Gravitation, University of Portsmouth, Dennis Sciama Building, Portsmouth, PO1 3FX, UK}
\affiliation{Institute of Space Sciences, ICE-CSIC, Campus UAB, Carrer de Can Magrans s/n, 08913 Bellaterra, Barcelona, Spain}

\author{S.~Gontcho A Gontcho\orcidlink{0000-0003-3142-233X}}
\affiliation{Lawrence Berkeley National Laboratory, 1 Cyclotron Road, Berkeley, CA 94720, USA}

\author{G.~Gutierrez}
\affiliation{Fermi National Accelerator Laboratory, PO Box 500, Batavia, IL 60510, USA}

\author{J.~Guy\orcidlink{0000-0001-9822-6793}}
\affiliation{Lawrence Berkeley National Laboratory, 1 Cyclotron Road, Berkeley, CA 94720, USA}

\author{C.~Hahn\orcidlink{0000-0003-1197-0902}}
\affiliation{Steward Observatory, University of Arizona, 933 N. Cherry Avenue, Tucson, AZ 85721, USA}

\author{H.~K.~Herrera-Alcantar\orcidlink{0000-0002-9136-9609}}
\affiliation{Institut d'Astrophysique de Paris. 98 bis boulevard Arago. 75014 Paris, France}
\affiliation{IRFU, CEA, Universit\'{e} Paris-Saclay, F-91191 Gif-sur-Yvette, France}

\author{K.~Honscheid\orcidlink{0000-0002-6550-2023}}
\affiliation{Center for Cosmology and AstroParticle Physics, The Ohio State University, 191 West Woodruff Avenue, Columbus, OH 43210, USA}
\affiliation{Department of Physics, The Ohio State University, 191 West Woodruff Avenue, Columbus, OH 43210, USA}
\affiliation{The Ohio State University, Columbus, 43210 OH, USA}

\author{M.~Ishak\orcidlink{0000-0002-6024-466X}}
\affiliation{Department of Physics, The University of Texas at Dallas, 800 W. Campbell Rd., Richardson, TX 75080, USA}

\author{D.~Kirkby\orcidlink{0000-0002-8828-5463}}
\affiliation{Department of Physics and Astronomy, University of California, Irvine, 92697, USA}

\author{T.~Kisner\orcidlink{0000-0003-3510-7134}}
\affiliation{Lawrence Berkeley National Laboratory, 1 Cyclotron Road, Berkeley, CA 94720, USA}

\author{A.~Kremin\orcidlink{0000-0001-6356-7424}}
\affiliation{Lawrence Berkeley National Laboratory, 1 Cyclotron Road, Berkeley, CA 94720, USA}

\author{M.~Landriau\orcidlink{0000-0003-1838-8528}}
\affiliation{Lawrence Berkeley National Laboratory, 1 Cyclotron Road, Berkeley, CA 94720, USA}

\author{L.~Le~Guillou\orcidlink{0000-0001-7178-8868}}
\affiliation{Sorbonne Universit\'{e}, CNRS/IN2P3, Laboratoire de Physique Nucl\'{e}aire et de Hautes Energies (LPNHE), FR-75005 Paris, France}

\author{M.~E.~Levi\orcidlink{0000-0003-1887-1018}}
\affiliation{Lawrence Berkeley National Laboratory, 1 Cyclotron Road, Berkeley, CA 94720, USA}

\author{R.~Miquel\orcidlink{}}
\affiliation{Instituci\'{o} Catalana de Recerca i Estudis Avan\c{c}ats, Passeig de Llu\'{\i}s Companys, 23, 08010 Barcelona, Spain}
\affiliation{Institut de F\'{i}sica d’Altes Energies (IFAE), The Barcelona Institute of Science and Technology, Edifici Cn, Campus UAB, 08193, Bellaterra (Barcelona), Spain}

\author{J.~Moustakas\orcidlink{0000-0002-2733-4559}}
\affiliation{Department of Physics and Astronomy, Siena College, 515 Loudon Road, Loudonville, NY 12211, USA}

\author{G.~Niz\orcidlink{0000-0002-1544-8946}}
\affiliation{Departamento de F\'{\i}sica, DCI-Campus Le\'{o}n, Universidad de Guanajuato, Loma del Bosque 103, Le\'{o}n, Guanajuato C.~P.~37150, M\'{e}xico}
\affiliation{Instituto Avanzado de Cosmolog\'{\i}a A.~C., San Marcos 11 - Atenas 202. Magdalena Contreras. Ciudad de M\'{e}xico C.~P.~10720, M\'{e}xico}

\author{W.~J.~Percival\orcidlink{0000-0002-0644-5727}}
\affiliation{Department of Physics and Astronomy, University of Waterloo, 200 University Ave W, Waterloo, ON N2L 3G1, Canada}
\affiliation{Perimeter Institute for Theoretical Physics, 31 Caroline St. North, Waterloo, ON N2L 2Y5, Canada}
\affiliation{Waterloo Centre for Astrophysics, University of Waterloo, 200 University Ave W, Waterloo, ON N2L 3G1, Canada}

\author{F.~Prada\orcidlink{0000-0001-7145-8674}}
\affiliation{Instituto de Astrof\'{i}sica de Andaluc\'{i}a (CSIC), Glorieta de la Astronom\'{i}a, s/n, E-18008 Granada, Spain}

\author{I.~P\'erez-R\`afols\orcidlink{0000-0001-6979-0125}}
\affiliation{Departament de F\'isica, EEBE, Universitat Polit\`ecnica de Catalunya, c/Eduard Maristany 10, 08930 Barcelona, Spain}

\author{A.~J.~Ross\orcidlink{0000-0002-7522-9083}}
\affiliation{Center for Cosmology and AstroParticle Physics, The Ohio State University, 191 West Woodruff Avenue, Columbus, OH 43210, USA}
\affiliation{Department of Astronomy, The Ohio State University, 4055 McPherson Laboratory, 140 W 18th Avenue, Columbus, OH 43210, USA}
\affiliation{The Ohio State University, Columbus, 43210 OH, USA}

\author{G.~Rossi}
\affiliation{Department of Physics and Astronomy, Sejong University, 209 Neungdong-ro, Gwangjin-gu, Seoul 05006, Republic of Korea}

\author{E.~Sanchez\orcidlink{0000-0002-9646-8198}}
\affiliation{CIEMAT, Avenida Complutense 40, E-28040 Madrid, Spain}

\author{D.~Schlegel}
\affiliation{Lawrence Berkeley National Laboratory, 1 Cyclotron Road, Berkeley, CA 94720, USA}

\author{H.~Seo\orcidlink{0000-0002-6588-3508}}
\affiliation{Department of Physics \& Astronomy, Ohio University, 139 University Terrace, Athens, OH 45701, USA}

\author{D.~Sprayberry}
\affiliation{NSF NOIRLab, 950 N. Cherry Ave., Tucson, AZ 85719, USA}

\author{M.~Walther\orcidlink{0000-0002-1748-3745}}
\affiliation{Excellence Cluster ORIGINS, Boltzmannstrasse 2, D-85748 Garching, Germany}
\affiliation{University Observatory, Faculty of Physics, Ludwig-Maximilians-Universit\"{a}t, Scheinerstr. 1, 81677 M\"{u}nchen, Germany}

\author{B.~A.~Weaver}
\affiliation{NSF NOIRLab, 950 N. Cherry Ave., Tucson, AZ 85719, USA}

\begin{abstract}
Recently the Dark Energy Spectroscopic Instrument (DESI) provided constraints on the expansion history from their Data Release 2 (DR2).  The DESI baryon acoustic oscillation (BAO) measurements are well described by a flat $\Lambda$CDM model, but the preferred parameters are in mild ($2.3\sigma$) tension with those determined from the cosmic microwave background (CMB).  The DESI collaboration has already explored a variety of solutions to this tension relying on variations in the late-time evolution of dark energy.  Here we test an alternative -- the introduction of an ``early dark energy'' (EDE) component.  We find that EDE models can alleviate the tension, though they lead to differences in other cosmological parameters that have observational implications.  Particularly the EDE models that fit the acoustic datasets prefer lower $\Omega_m$, higher $H_0$, $n_s$ and $\sigma_8$ in contrast to the late-time solutions.  We discuss the current status and near-future prospects for distinguishing amongst these solutions.
\end{abstract}

\maketitle
\flushbottom

\section{Introduction}

The measurement of the distance-redshift relation through baryon acoustic oscillations (BAO) in the second data release (DR2) of the Dark Energy Spectroscopic Instrument (DESI) collaboration \cite{Snowmass2013.Levi,DESICollaboration2016,DESICollaboration2016a,DESI2022.KP1.Instr} has revealed a mild tension between acoustic waves measured in the cosmic microwave background (CMB) radiation and in BAO \cite{DESI.DR2.BAO.cosmo} when interpreted within the framework of $\Lambda$CDM.  This tension has persisted for many years, over several experiments, and has been growing in significance (e.g.\ compare Fig.~1 to Fig.~12 of \cite{Planck18-VI}).  It is particularly puzzling, because BAO rely on a characteristic scale in the clustering of galaxies that arises from acoustic waves propagating in the coupled baryon-photon fluid in the pre-recombination Universe (for a recent review see \cite{KP4s2-Chen}, for textbook treatments see e.g.\ \cite{Dodelson20,Huterer23}).  This is the exact same physics as gives rise to the anisotropies in the CMB that have been exquisitely measured by WMAP \cite{WMAP9}, Planck \cite{Planck18-I} and a host of ground-based experiments, most recently ACT \cite{Madhavacheril24,Qu24,MacCrann24} and SPO \cite{SPT}.

\begin{figure}
    \centering
    \resizebox{\columnwidth}{!}{\includegraphics[scale=1]{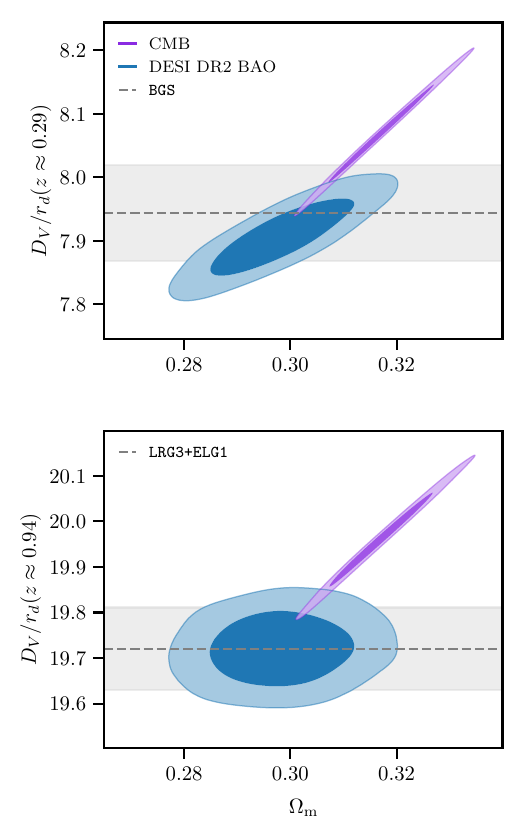}}
    \caption{The 68\% and 95\% marginal posteriors for $\Omega_m$ and $D_V/r_d$ at $z\simeq 0.3$ and $z\simeq 0.9$ from the CMB (purple) and BAO (blue) datasets are described in the text, within the context of $\Lambda$CDM.  For the CMB we use the combination of Planck PR4 and Planck+ACT lensing, while for the BAO we use DESI DR2 \cite{DESI.DR2.BAO.cosmo}.  The horizontal dashed line with shaded band is the central value and $1\,\sigma$ error measured in DR2 for \texttt{BGS} at $z=0.295$ (top) and \texttt{LRG3+ELG1} at $z=0.934$ (bottom) for reference. The $\Lambda$CDM fit to the full DR2 dataset predicts distances in good agreement with the measured values at these redshifts while that fit to the CMB overpredicts these distances.}
    \label{fig:OmM-DVrd}
\end{figure}

Individually the acoustic signals measured in the CMB and in DESI are each consistent with $\Lambda$CDM \cite{WMAP9,Planck18-I,DESI.DR2.BAO.cosmo}, however there is a $2.3\,\sigma$ tension\footnote{If we include CMB lensing.  If not, the `tension' is $2.0\,\sigma$ \cite{DESI.DR2.BAO.cosmo}.} in the values of the cosmological parameters allowed by the two datasets \cite{DESI.DR2.BAO.cosmo} -- see Fig.~\ref{fig:OmM-DVrd}.  If we assume the tension is not due to a statistical fluctuation or systematics in the measurements, then it can be resolved by changes to the model at low redshift (e.g.\ evolving dark energy) or between $z\sim 10^4$ and $10^3$ (e.g.\ ``early'' dark energy). Several late time solutions are explored in refs.~\cite{DESI.DR2.BAO.cosmo,Y3.cpe-s1.Lodha.2025,Y3.cpe-s2.Elbers.2025}.  In this paper we discuss a solution based on modifying the model at early times.
 
Specifically we study whether including an `early dark energy' (EDE) component that contributes to the expansion rate between matter-radiation equality and recombination can improve the agreement between high- and low-$z$ measures of acoustic oscillations (i.e.\ the CMB and BAO).  The EDE component causes an increase in $H(z)$ at early times, decreasing the acoustic scale.  The adjustments in the other parameters (primarily an increase in $H_0$ and slight lowering of $\Omega_m$) to hold the well-measured acoustic scale of the CMB fixed allow a better fit with the DESI data.

The outline of this paper is as follows.  In the next section (\S\ref{sec:data}) we describe the DESI data and the likelihoods that we employ.  This is followed (\S\ref{sec:EDE}) by a discussion of the scalar field model that we use as an example of a modification of the expansion history at early times.  Our fits and basic results are presented in \S\ref{sec:results}.  We then discuss the implications of these results for future measurements, and routes towards distinguishing between early- and late-time solutions, in \S\ref{sec:future}.  Finally we conclude in \S\ref{sec:conclusions}.

\section{Data and cosmological inference}
\label{sec:data}

Our focus shall be on the two datasets that use the acoustic signature to provide cosmological constraints, specifically the CMB (from Planck and ACT) and BAO (from DESI).  One of the advantages of the EDE model we shall discuss is that only the robust acoustic data are needed for data-driven (rather than prior dominated) constraints on the model.  This is in contrast to solutions that modify the DE evolution at late times ($w_0w_a$ model), where additional data (e.g.\ Type Ia SNe) are necessary.

We use the BAO measurements of the transverse ($D_M$) and line-of-sight ($D_H$) distances to 7 redshift slices from DESI DR2 \cite{DESI.DR2.DR2}.  We follow common convention and define the combinations
\begin{align}
    D_V(z)    &= \left[ z D_M^2(z)D_H(z)\right]^{1/3} \\
    F_{AP}(z) &= D_M(z) / D_H(z)
\end{align}
which measure the isotropically averaged distance to, and the anisotropy of the distances at, redshift $z$.

The DESI Collaboration used a robotic, 5000-fiber spectrograph \cite{DESI2022.KP1.Instr,FocalPlane.Silber.2023,FiberSystem.Poppett.2024,Corrector.Miller.2023} on the 4-m Nicholas U. Mayall Telescope at the Kitt Peak National Observatory to measure the redshifts \cite{Spectro.Pipeline.Guy.2023,Redrock.Bailey.2024} of galaxies and quasars with the goal of constraining cosmology via inhomogeneities in both galaxy density and the intergalactic medium \cite{DESICollaboration2016}.  In its first three years of operation DESI measured redshifts for over 30 million galaxies and quasars \cite{TS.Pipeline.Myers.2023,SurveyOps.Schlafly.2023}.  The DR2 dataset that we focus on, containing data taken between 14 May 2021 and 9 April 2024, uses $\sim 14$ million of these \cite{DESI.DR2.DR2}, plus the Ly$\alpha$ forest in the spectra of 820,000 QSOs \cite{DESI.DR2.BAO.lya}, to measure BAO in 7 redshift slices \cite{Y3.clust-s1.Andrade.2025,DESI.DR2.BAO.cosmo} spanning $0.1<z<3.5$.  Further information on the data set, and the BAO methodology --- which largely follows that used in DR1 \cite{DESI2024.III.KP4,DESI2024.IV.KP6,DESI2024.VI.KP7A,KP4s2-Chen} --- can be found in refs.~\cite{DESI.DR2.BAO.cosmo,Y3.clust-s1.Andrade.2025,DESI.DR2.BAO.lya}.

Inclusion of the full information from the CMB is critical to constraining a model such as EDE.  We follow ref.~\cite{DESI.DR2.BAO.cosmo} in using the temperature ($TT$), polarization ($EE$) and cross ($TE$) power spectra from Planck, specifically using the \texttt{simall}, \texttt{Commander} (for $\ell<30$; \cite{Planck18-V}) and \texttt{CamSpec} (for $\ell\geq30$; \cite{Rosenberg22}) likelihoods based upon the latest Planck release (NPIPE maps \cite{PR4-NPIPE}).
In addition to the primary anisotropies, we use the combination of Planck and ACT DR6 CMB lensing detailed in ref.~\cite{Madhavacheril24}.  While there are several comparable choices of CMB likelihood, inferences about EDE are relatively insensitive to these choices \cite{McDonough24,Efstathiou24} so we choose to follow those in ref.~\cite{DESI.DR2.BAO.cosmo}.

For some of our results we shall also make use of SNe to constrain the low-$z$ distance-redshift relation.  We have chosen the Union3 sample \cite{Rubin23} for illustrative purposes, and the choice of the SNe data set does not impact our finding on EDE.  In such cases we use `uncalibrated' SNe distances, i.e.\ marginalizing over an unknown absolute magnitude.  The use of `calibrated' SNe for constraining $H_0$ will be discussed later.

In the following, our cosmological inferences will be performed using \texttt{desilike}\footnote{Publicly available: \url{https://github.com/cosmodesi/desilike}}. The posterior profiling is performed through the \texttt{iminuit} \cite{iminuit} minimiser\footnote{\url{https://github.com/cosmodesi/desilike/blob/main/desilike/profilers/minuit.py}}, the Monte Carlo Markov chains (MCMC) use the \texttt{emcee} \cite{Foreman-Mackey2013} sampler\footnote{\url{https://github.com/cosmodesi/desilike/blob/main/desilike/samplers/emcee.py}}, and we use \texttt{GetDist} \cite{getdist} to display the posteriors. The priors assumed in our analysis are given in Table~\ref{tab:priors}.
For the late-time dark energy (i.e.\ $w_0w_a$) model we make use of the posteriors and the chains from ref.~\cite{DESI.DR2.BAO.cosmo}, where they are described in some detail.

\begin{table}
\centering
\begin{tabular}{l|c}
\toprule
Parameter        & Prior \\ 
\midrule
$\Omega_m$       & $\mathcal{U}[0.1, 0.9]$ \\
$\omega_b$       & $\mathcal{U}[0.021, 0.025]$ \\
$H_0$            & $\mathcal{U}[20,100]$ \\
$\mathrm{ln}(10^{10}A_s)$        & $\mathcal{U}[2.7, 3.3]$ \\
$n_s$            & $\mathcal{U}[0.9, 1.04]$ \\
$\tau_{\mathrm{reio}}$            & $\mathcal{U}[0.03, 0.1]$ \\
\hline
$f_{\rm EDE}$        & $\mathcal{U}[0,0.3]$ \\
$\log_{10}(a_c)$ & $\mathcal{N}(-3.531, 0.1)$ \\
$\Theta_{\mathrm{ini}}$ & $\mathcal{U}[0, \pi]$ \\
\bottomrule
\end{tabular}
\caption{The priors used in our analysis.  The parameter names have their usual meanings, or are defined in the text.  The notation $\mathcal{U}[a,b]$ indicates a uniform prior on the closed interval $[a,b]$, while $\mathcal{N}(\mu,\sigma)$ indicates a normal distribution with mean $\mu$ and variance $\sigma^2$. We use the standard priors for the different nuisance parameters in the CMB likelihoods.}
\label{tab:priors}
\end{table}

\section{Early dark energy}
\label{sec:EDE}

Fig.~\ref{fig:OmM-DVrd} shows the 68\% and 95\% marginal posteriors for $\Omega_m$ and $D_V/r_d$ at $z\simeq 0.29$ and $z\simeq 0.94$ from the CMB and BAO datasets described above.  For the CMB the inference assumes the $\Lambda$CDM model, and the same is true of the BAO data but since this is very close to what the BAO natively constrain the model dependence of this measurement is much weaker.  Each dataset is individually consistent with $\Lambda$CDM but within the $\Lambda$CDM paradigm the parameters preferred by each dataset are in (weak, $2.3\,\sigma$) tension \cite{DESI.DR2.BAO.cosmo}.  The CMB data prefer a larger distance and a higher value of $\Omega_m$ than do the BAO data.  The same distance overprediction holds for all of the $z<1$ distances constrained by DESI. Interestingly this is the same direction as the offset predicted by the EDE model, so we expect EDE to alleviate some of this tension.

The specific example of an early-time modification to the expansion history that we shall study is the EDE model.  This is usually realized as a canonically normalized, minimally coupled scalar field moving in a potential
\begin{equation}
  V(\phi) = V_0 \left[ 1 - \cos\frac{\phi}{f} \right]^n \quad , \quad V_0\equiv m^2f^2
\end{equation}
where $f\sim M_{\rm Pl}$ is a decay constant, $m\sim 10^{-28}\,$eV a mass and $n\approx 3$ for the model to be viable.  We shall follow this practice as well, though we expect the arguments to be more general, and we shall fix $n=3$ throughout. Rather than use $m$ and $f$ as our parameters, we shall follow the usual convention \cite{McDonough24} and work with variables more closely related to the observations: $\log_{10}(a_c)$ and $f_{\rm EDE}$.  These are respectively the (logarithm of the) scale factor where the EDE density peaks and the fraction of $\rho_{\rm crit}$ that it makes up at that time.  We will see that $a_c\sim 10^{-3.5}$ and $f\sim 5-10\%$ will allow us to simultaneously fit the acoustic signatures in the CMB and DESI data with a single model.  The final `additional' parameter in the EDE model is the initial value of the scalar field, $\phi_i$, which we will find prefers to be $\Theta_i\equiv \phi_i/f\approx\pi$.  Note that in such situations the field evolves over the full potential range during cosmic evolution, in contrast to many scalar-field-based models of `late time' dark energy for which the field evolves very little.

The EDE model was originally proposed \cite{Karwal16,Poulin2019,Smith2020} to resolve the Hubble tension \cite{Knox20,Abdalla22}, but our goal is to use it as an example of a model that modifies the expansion history before recombination, and hence the sound horizon or `normalization' of the BAO ruler.  A recent review of EDE models, and a comparison against current observations, can be found in refs~\cite{Poulin23,McDonough24}, while \cite{Jiang25} included the latest DESI DR1 BAO data.

The dynamics of the scalar field, and its fluctuations, can be straightforwardly computed \cite{Kodama84,Caldwell98,Poulin2019}, and we use the implementation of \texttt{AxiCLASS}\footnote{We use \texttt{AxiCLASS} (\url{https://github.com/PoulinV/AxiCLASS}) via the python wrapper: \url{https://github.com/cosmodesi/cosmoprimo}.} \cite{Blas2011,Poulin2018,Smith2020}, based on \texttt{CLASS} \cite{Lesgourgues11}, to compute the background evolution and linear perturbations. To speed up the cosmological inference, we decide to emulate the required outputs of \texttt{AxiCLASS} with a neural network\footnote{We use a multilayer perceptron whose implementation is available here: \url{https://github.com/cosmodesi/cosmoprimo/blob/main/cosmoprimo/emulators/tools/mlp.py}.}.

\begin{figure}
    \centering
    \hspace{-0.5cm}\includegraphics[scale=1]{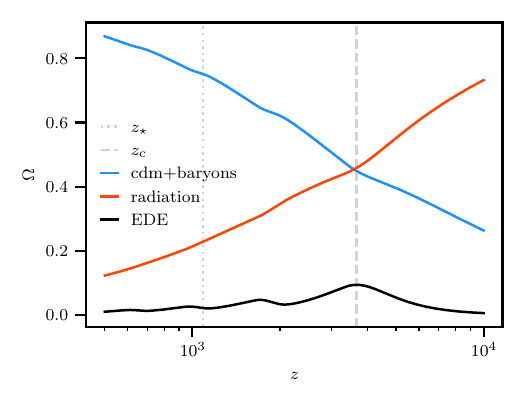}
    \caption{Evolution of the energy densities in radiation (orange), matter (blue) and EDE (black) with redshift for our best-fitting EDE model, expressed as a fraction of the critical density.  The EDE density peaks ($z_{\rm c}$) near matter radiation equality ($z_{\rm eq}$) with an amplitude of $9.3\%$ and is almost entirely gone by recombination ($z_{\star}$).}
    \label{fig:fede_vs_z}
\end{figure}

At early times $\phi$ is pinned by Hubble drag.  It begins to act like a dark energy component when $V(\phi)\sim H^2$, at which point the field begins to roll down the potential and EDE becomes dynamical.  As the field oscillates about the minimum of the potential, $V(\phi)\propto\phi^{2n}$, the EDE dilutes as an effective fluid with equation of state, $w=(n-1)/(n+1)$, i.e.\ $\rho_{\rm EDE}\propto a^{-3(1+w)}\approx a^{-9/2}$.
\cref{fig:fede_vs_z} shows the evolution of $\rho_{\rm EDE}$ for the model that best fits the data described in \S\ref{sec:data}.

\begin{figure*}
    \centering
    \includegraphics[width=\textwidth]{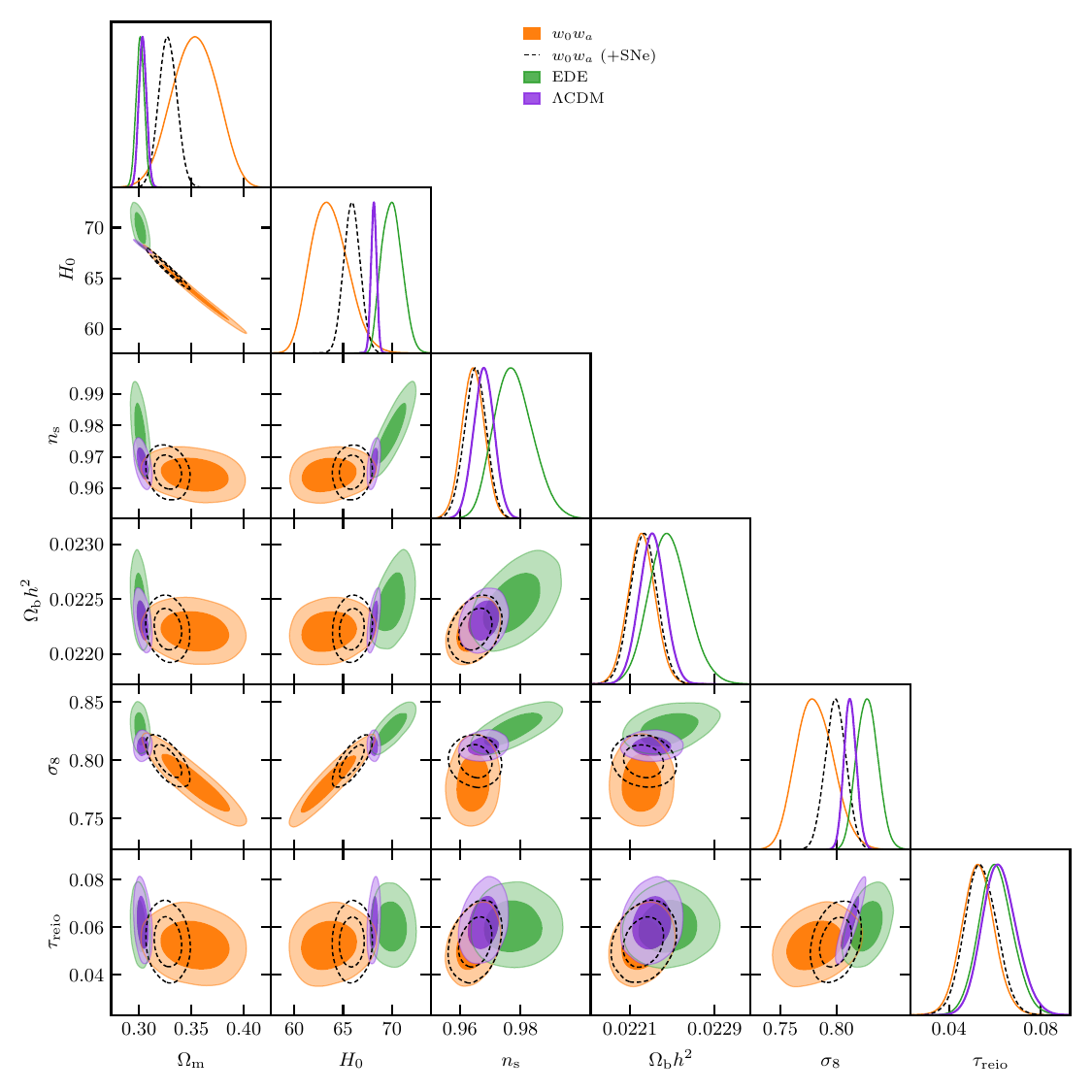}
    \caption{Marginal 68\% and 95\% posteriors for the cosmological parameters in common to all models, for the combination of CMB and BAO data. Purple, green, orange contours are for the $\Lambda$CDM, EDE and $w_0w_a$ models, respectively, while the black dashed contour is the $w_0w_a$ model including SNe (Unions3). Note the EDE model is consistent with higher values of $H_0$, and slightly smaller values of $\Omega_m$, than $\Lambda$CDM (as discussed in the text). The $w_0w_a$ model shows the opposite trend.}
    \label{fig:base-params}
\end{figure*}

\begin{figure*}
    \centering
    \includegraphics[width=\textwidth]{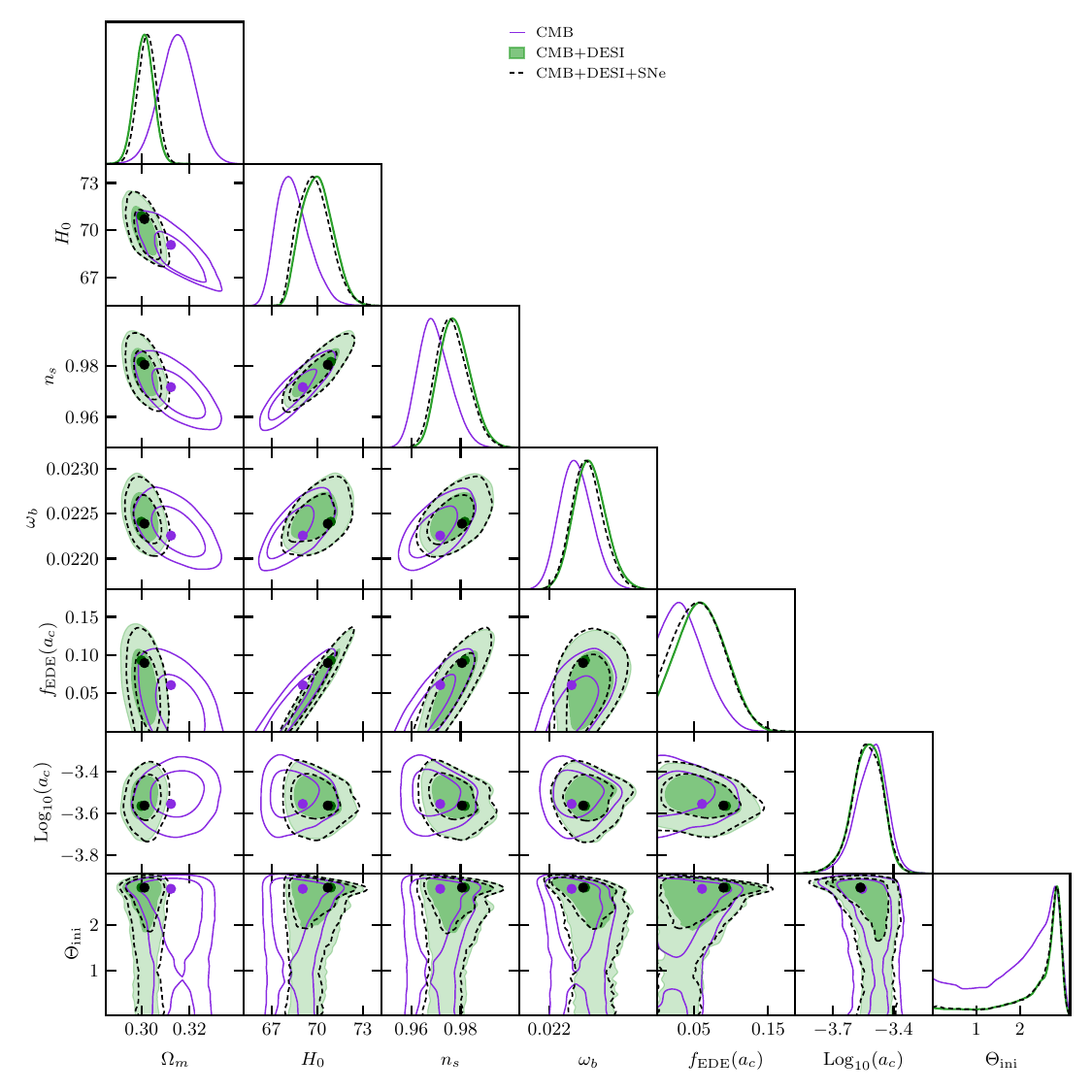}
    \caption{Marginal posteriors for select parameters of the EDE model, for different choices of data (CMB [purple], CMB+BAO [green] and CMB+BAO+SNe [black]).  Note that the model is well constrained by the CMB, but the parameters shift when the CMB data are combined with DESI BAO.  The parameters remain stable under the addition of SNe data (Union3).  When low- and high-$z$ data are included the model prefers lower $\Omega_m$ and higher $H_0$ than with CMB alone. The solid dots are the MAP for the different choices of data, indicating no serious parameter projection effects for any data combination.}
    \label{fig:ede-params}
\end{figure*}

The introduction of EDE allows us to hold the acoustic scale
\begin{equation}
  \theta_\star = \frac{r_s}{\chi_\star} \quad ; \quad
  r_s = \int_{z_\star}^\infty\frac{c_s\,dz}{H(z)} \quad , \quad
  \chi_\star = \int_0^{z_\star}\frac{dz}{H(z)}
\end{equation}
fixed, while varying $H(z)$ at both early and late times.  In the above $z_{\star}\approx 10^3$ is the redshift of last scattering.  An increase in $H(z>z_\star)$, from EDE, leads to a decrease in\footnote{Although we shall distinguish between the sound horizon at recombination ($r_s$) and at the drag epoch ($r_d$) throughout, in fact for the models of interest $r_d/r_s\approx 1.015$ is very close to constant.} $r_s$.  Increasing $H(z<z_\star)$, for example by increasing $\omega_m$ or $h$, can compensate by decreasing $\chi_\star$ to leave $\theta_\star$ fixed, as required by the tight constraints from the CMB.  The field is required to decay quickly so that the diffusion damping scale\footnote{Whereas the CMB acoustic scale, $r_s$, and the BAO acoustic scale, $r_d$, depend linearly on $t_\star$, the diffusion damping scale depends upon $\sqrt{t_\star}$.  The combination of the two thus allows a constraint on early-time dynamics \cite{Hu96}.} in the CMB \cite{Silk68} is not excessively modified.

\section{Results}
\label{sec:results}

Figure \ref{fig:base-params} shows the marginal likelihoods of several key cosmological parameters for the $\Lambda$CDM, EDE and $w_0w_a$ models when fit to the combination of CMB and BAO datasets.  Best fits and $1\,\sigma$ confidence levels are given in Tables \ref{tab:param_constraints} and \ref{tab:param_constraints_lcdm}.  Both modifications to late time DE dynamics and EDE induce shifts in the best-fit cosmological parameters (compared to $\Lambda$CDM) in order to continue to fit the observations.  In both cases the physical matter density ($\omega_m$) increases while $\Omega_m$ decreases and $H_0$ increases for EDE and the opposite happens for $w_0w_a$.

For EDE $\omega_m$ increases due to the need to compensate the EDE dynamics at $z\sim z_c$ while holding $\theta_\star$ fixed.  Since the scalar field is `stiff', it contributes to the background expansion ($H$) but not to the potentials inside the horizon.  It thus causes sub-horizon modes to grow more slowly near matter-radiation equality and recombination.  This can be compensated by an increase in $\omega_m$ in order to counteract the enhancement of the first acoustic peak in the CMB \cite{Lin19,Hill2020,Vagnozzi21}.  To fix the spectrum at higher $\ell$ the spectral index ($n_s$) also needs to increase, and there is a small increase in the normalization ($A_s$) \cite{McDonough24,Efstathiou24} as shown in Fig.~\ref{fig:ede-params}.  The increase in $\omega_m$ and $h$ increase the expansion rate at low redshift, since $H(z)\propto\sqrt{h^2+\omega_m e(z)}=h\sqrt{1+\Omega_m e(z)}$ with $e(z)=(1+z)^3-1$ for $z\ll 100$.  This causes the required decrease in $\chi_\star$ so that $\theta_\star=r_s/\chi_\star$ is left unchanged even though $r_s$ is decreased by 3\% (measured in Mpc).  We find $H(z)$ is $\approx 5$\% higher over the range $0<z<1$ in the best-fitting EDE model compared to the best-fitting $\Lambda$CDM model.  The distances to $z\ll 1$ are thus lowered modestly because $H_0$ has increased.  Since the majority of the change is through $H_0$, if distances are measured in $h^{-1}$Mpc then they change minimally.  The quantity $H_0r_d$, being the BAO analogue of $\theta_\star$ in the CMB, is well constrained by BAO and thus $r_d$ is also almost unchanged in $h^{-1}$Mpc units.

For $w_0w_a$ only the late time dynamics is altered.  Thus the drag scale ($r_d$) is unchanged and the shift in $\omega_m$ occurs so as to hold $\chi_\star$, and hence $\theta_\star$, fixed when the DE evolution at $z<1$ is modified.  The best-fitting model requires the dark energy to evolve quickly (in a fraction of the Hubble time) at late times, leading to a change in $\chi(z)$ at $z<1$ \cite{DESI.DR2.BAO.cosmo}.  In contrast to EDE, the late-time solution has a modest decrease in $n_s$, $A_s$ and $\sigma_8$.

While the constraints in Fig.~\ref{fig:base-params} are superficially similar, the isotropic BAO distance ($D_V/r_d$) below $z\simeq 1$ is smaller in the EDE model than the fiducial $\Lambda$CDM model fit to Planck, in better agreement with the DESI BAO data.  By contrast $F^{AP}=D_M/D_H$ is almost unchanged by the introduction of EDE, rather than following the data to increase towards lower $z$.  However the DESI constraints on $F^{AP}$ are less stringent than on $D_V/r_d$, so this mismatch is not significant.
The $\Lambda$CDM and EDE cases provide better agreement with the high-$z$ data (i.e.\ the Ly$\alpha$ points), while the $w_0w_a$ case does less well.  The errors at high $z$ are still sufficiently large, however, that this discrepancy is statistically insignificant.

\begin{figure}
    \centering
    \includegraphics[scale=1]{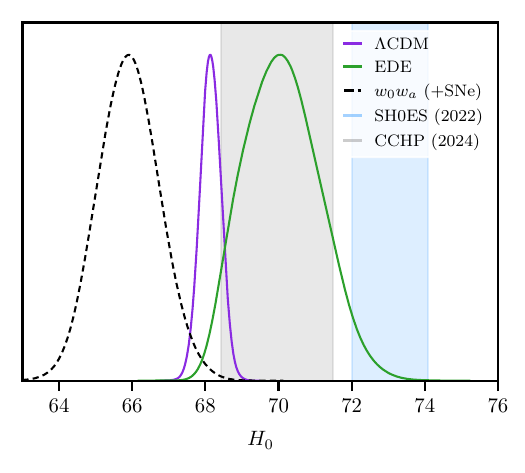}
    \caption{The $H_0$ posteriors for our three models fit to the combination of CMB+BAO data, including +SNe (Union3) for the $w_0w_a$ model.  The values inferred from the local distance ladder by the SH0ES and CCHP teams are shown as vertical bands.}
    \label{fig:H0}
\end{figure}

Figure \ref{fig:H0} shows the marginal posterior for $H_0$ for our three models, fit to the acoustic datasets (CMB+BAO) for $\Lambda$CDM and EDE and with the inclusion of (uncalibrated) SNe from Union3 for the $w_0w_a$ model.  The inferences from the distance ladder by the SH0ES \cite{SH0ES22} and CCHP \cite{CCHP24} teams are shown as vertical bands.  As has been noted previously \cite{Poulin23,McDonough24,Efstathiou24} EDE models consistent with CMB have their most-probable values below the SH0ES result on $H_0$, though the tension is lessened compared to $\Lambda$CDM.  The $w_0w_a$ models fare even more poorly in this regard.  However the CCHP result is consistent with the predictions of all three of our models, though only marginally for $w_0w_a$.  Since the $H_0$ value allowed by the combination of CMB and BAO within the EDE model is higher than within $\Lambda$CDM it overlaps considerably with that inferred by CCHP.
If we include the $H_0$ values in our fits then the preference for EDE over $\Lambda$CDM or $w_0w_a$ improves (see Table \ref{tab:delta-chi2}).  New data and improved analysis methods \cite{Riess24,Hoyt25} are already reducing the systematic errors in the local distance scale, and will help to resolve this discrepancy between the two $H_0$ measures.

To compare the different models, we show\footnote{We use our emulators to find the best fits shown in Table \ref{tab:param_constraints}, but then compute $\chi^2$ directly, i.e.\ without the emulator, at the best fit cosmology.} in Table~\ref{tab:delta-chi2} the $\chi^2_{\rm MAP}$ at the best fit parameters.  The best-fit for $w_0w_a$ is taken from ref.~\cite{DESI.DR2.BAO.cosmo}. Note that we found a slightly larger $\Delta\chi^2_{\rm MAP}$ for $w_0w_a$ over $\Lambda$CDM than quoted in ref.~\cite{DESI.DR2.BAO.cosmo}. This is because we are computing the $\chi^2_{\rm MAP}$ for $\Lambda$CDM at the best fit found by our emulator instead of the best fit found in \cite{DESI.DR2.BAO.cosmo}, resulting to a difference about $\Delta \chi^2 \sim 1$. Finally, we speculate that the residual small mismatch between our computation using \texttt{CLASS} and that of ref.~\cite{DESI.DR2.BAO.cosmo} using \texttt{CAMB} may be resolved by increasing the accuracy settings in \texttt{CLASS}.

Regardless of these details, we find that, for CMB+BAO, the addition of EDE improves the fit over the $\Lambda$CDM model by $\Delta\chi^2_{\rm MAP}=7.4$, close to a $2\sigma$ improvement given the 3 additional degrees of freedom. As expected, the inclusion of (uncalibrated) SNe data does not further increase $\Delta\chi^2_{\rm MAP}$ as the EDE model does not match the evolution desired by the SNe.  By comparison, for the same data, late time evolving dark energy improves the $\chi^2_{\rm MAP}$ by 13.2 using only two new parameters.  This increases to 19 when we include the SNe data, since a rapidly evolving late-time dark energy can better fit the SNe data below $z\simeq 0.5$.  If we add measures of $H_0$ from the local distance scale, then the relative preference for $w_0w_a$ over EDE is reduced (see Table \ref{tab:delta-chi2}), since the $w_0w_a$ models reduce $H_0$ while the EDE models increase it.

\begin{figure}
    \centering
    \hspace{-1cm}\includegraphics[scale=1]{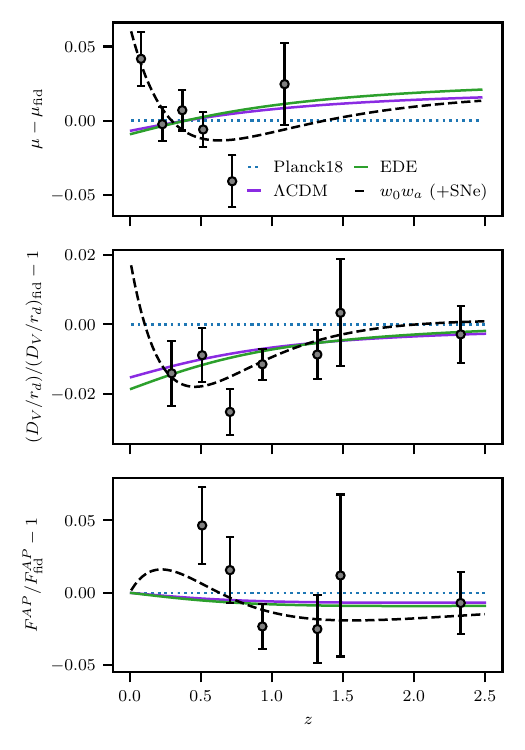}
    \caption{Predicted low-$z$ expansion history for the best-fit models.  In each panel the purple, green and black lines are the predictions of the $\Lambda$CDM, EDE and $w_0wa$ models that best fit the combination of CMB+BAO data discussed in the text, compared to the DESI `fiducial' $\Lambda$CDM model (also used as the fiducial cosmology in ref.~\cite{DESI.DR2.BAO.cosmo}).  The top panel shows the distance modulus, such as would be measured by SNe, the middle panel the isotropic distance scale constrained by BAO and the lower panel the distance ratio.  The points with error bars in the top panel are binned data from the Union3 SNe dataset, with the weighted mean set to zero to account for the unknown absolute magnitude.  The points in the middle and lower panel are from DESI DR2. Note that $\Lambda$CDM and EDE predict similar late-time expansion, but the $\Lambda$CDM model provides a worse fit to the data combination than the EDE model (Table \ref{tab:delta-chi2}).}
    \label{fig:data_vs_model}
\end{figure}

\begin{table*}
    \centering
    \caption{Constraints on the parameters of the different models for each of the data combinations considered in this analysis. The central values are the best fit value from the \texttt{iminuit} minimization while the errors are the $1\sigma$ credible interval from the corresponding chains. The last column shows the $\Delta\chi^2$ difference of the MAP points for each model. For $\Lambda$CDM and EDE models, the $\chi2$ is computed without the emulator at the best fit found with the emulator. Additional parameters are given in Table~\ref{tab:param_constraints_lcdm}}
    \label{tab:param_constraints}
    \begin{adjustbox}{width=1\textwidth}
    \begin{tabular}{l|ccccccccccc}
        \toprule
Data (Model)              &  $\Omega_m$         & $H_0$            & $\sigma_8$          & $f_{\mathrm{EDE}}$ & $\log_{10}(a_c)$ & $w_0$ & $w_a$ \\
\midrule
CMB ($\Lambda$CDM)         & $0.3188 \pm 0.0067$ & $67.02 \pm 0.48$ & $0.8122 \pm 0.0052$ & - & - & - & - \\
\midrule 
CMB+DESI ($\Lambda$CDM)    & $0.3037 \pm 0.0037$ & $68.12 \pm 0.28$ & $0.8101 \pm 0.0055$ & - & - & - & - \\
CMB+DESI (EDE)             & $0.2999 \pm 0.0038$ & $70.9  \pm 1.0$  & $0.8283 \pm 0.0093$ & $0.093 \pm 0.031$ & $-3.564 \pm 0.075$ & - & - \\
CMB+DESI ($w_0w_a$)        & $0.353 \pm  0.021$  & $63.5  \pm 1.9$  & $0.780 \pm 0.016$   & - & - & $-0.42 \pm 0.21$ & $-1.75 \pm 0.58$ \\
\midrule 
CMB+DESI+SNe ($\Lambda$CDM)& $0.3047 \pm 0.0036$ & $68.04 \pm 0.28$ & $0.8102 \pm 0.0054 $& - & - & - & - \\
CMB+DESI+SNe (EDE)         & $0.3012 \pm 0.0037$ & $70.7  \pm 1.0$  & $0.8277 \pm 0.0097 $& $0.09 \pm 0.032$ & $-3.562 \pm 0.077$ & - & - \\
CMB+DESI+SNe ($w_0w_a$)    & $0.3270 \pm 0.0086$ & $65.92 \pm 0.84$ & $0.7989 \pm 0.0093$ & - & - & $-0.672 \pm 0.088$ & $-1.06 \pm 0.29$ \\
        \bottomrule
    \end{tabular}
    \end{adjustbox}
\end{table*}

\begin{table*}
    \centering
    \caption{As in Table~\ref{tab:param_constraints} but for the `base' $\Lambda$CDM parameters. The central values are the best fit value from the \texttt{iminuit} minimization while the errors are the $1\sigma$ credible interval from the corresponding chains.}
    \label{tab:param_constraints_lcdm}
    \begin{adjustbox}{width=1\textwidth}
    \begin{tabular}{l|cccccccc}
        \toprule
Data (Model)                   & $\omega_m$          & $H_0$            & $n_s$               & $\omega_b$            & $\ln\left(10^{10}A_s\right)$ & $\tau_{\mathrm{reio}}$ \\
\midrule
CMB ($\Lambda$CDM)         & $0.1432 \pm 0.0010$ & $67.02 \pm 0.48$ & $0.9593 \pm 0.0039$ & $0.02212 \pm 0.00013$ & $3.044 \pm 0.013$ & $0.0547 \pm 0.0073$ \\
\midrule 
CMB+DESI ($\Lambda$CDM)    & $0.1409 \pm 0.00061$ & $68.12 \pm 0.28$ & $0.9672 \pm 0.0034$ & $0.02229 \pm 0.00012$ & $3.056 \pm 0.014$ & $0.0621 \pm 0.0075$ \\
CMB+DESI (EDE)             & $0.1507 \pm 0.0035$ & $70.9 \pm 1.0$   & $0.9817 \pm 0.0063$ & $0.02241 \pm 0.00018$ & $3.067 \pm 0.017$ & $0.0582 \pm 0.0074$ \\
CMB+DESI ($w_0w_a$)        & $0.142 \pm 0.021$  & $63.5  \pm 1.9$  & $0.9632 \pm 0.0037$ & $0.02218 \pm 0.00013$ & $3.037 \pm 0.013$ & $0.0520 \pm 0.0071$ \\
\midrule 
CMB+DESI+SNe ($\Lambda$CDM)& $0.1410 \pm 0.00060$ & $68.04 \pm 0.28$ & $0.9668 \pm 0.0033$ & $0.02228 \pm 0.00012$ & $3.055 \pm 0.013$ & $0.0605 \pm 0.0073$ \\
CMB+DESI+SNe (EDE)         & $0.1505 \pm 0.0036$ & $70.7  \pm 1.0$  & $0.9806 \pm 0.0064$ & $0.02239 \pm 0.00018$ & $3.066 \pm 0.014$ & $0.0578 \pm 0.0073$ \\
CMB+DESI+SNe ($w_0w_a$)    & $0.1421 \pm 0.0086$ & $65.92 \pm 0.84$ & $0.9646 \pm 0.0036$ & $0.02221 \pm 0.00013$ & $3.039 \pm 0.013$ & $0.0529 \pm 0.0070$ \\
        \bottomrule
    \end{tabular}
    \end{adjustbox}
\end{table*}

\begin{table}
\centering
\caption{Relative goodness-of-fit of the models in Tables \ref{tab:param_constraints} and \ref{tab:param_constraints_lcdm} (and discussed in the text) for various dataset combinations.  For each data combination we quote the $\chi^2$ difference at the MAP point, $\Delta\chi^2_{MAP}$, between $\Lambda$CDM and the EDE model and between the $\Lambda$CDM and $w_0w_a$ model.  The term ``SNe'' refers to `uncalibrated' SNe to measure the distance-redshift relation while CCHP and SH0ES refer to the `calibrated' distance ladder (including SNe).  A positive number indicates a preference over $\Lambda$CDM.  Note that for simplicity when including CCHP or SH0ES we simlpy report $\Delta\chi^2_{MAP}$ for the models already described rather than refitting to the combined data. }
\begin{tabular}{l|cc}
\toprule
Data      &$\Lambda$CDM - EDE & $\Lambda$CDM - $w_0w_a$ \\ 
\midrule
CMB+BAO   & 7.4 & 13    \\
+SNe      & 7.5 & 19    \\
+CCHP     & 8.7 & -2.9  \\
+SH0ES    & 25  & -48.5 \\
+SNe+CCHP & 8.7 & 14    \\
+SNe+SH0ES& 26  & -4.6  \\
\bottomrule
\end{tabular}
\label{tab:delta-chi2}
\end{table}

\section{Implications for future measurements}
\label{sec:future}
Almost by design, the EDE model\footnote{We do not consider a combination of early- and late-time evolving dark energy in this paper.} shifts the sound horizon while leaving the shape of the late-time expansion history unchanged from $\Lambda$CDM. As Fig.~\ref{fig:data_vs_model} shows, this allows us to provide a good match to the BAO data while simultaneously matching the CMB.  However it does not provide as good a fit to the rapid increase in $\mu-\mu_{\rm fid}$ at low $z$ preferred by the `uncalibrated' SNe data (the case of Union3 \cite{Rubin23} is shown in Fig.~\ref{fig:data_vs_model}, the data from DES Y5 \cite{DES-Y5SNe} appear similar while the tendency is less pronounced in the Pantheon+ \cite{Scolnic21} dataset).  We expect further improvements to the SNe data with ongoing and upcoming surveys \cite{Rigault25,LSST09}, and improvements in the low-$z$ BAO measurements by extending the footprint of DESI; these measurements will be highly informative regarding the viability of an early-time solution to the current tension.  If the BAO, even used as an uncalibrated ruler, show a similar behavior to the SNe then an EDE solution will be (more) disfavored.

\begin{figure}
\centering
\resizebox{\columnwidth}{!}{\includegraphics[scale=1]{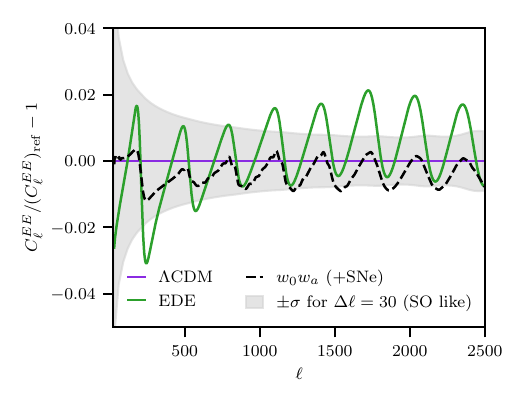}}
\caption{$E$-mode polarization auto-spectrum between the best fit EDE model for the combination of CMB+BAO data (green line) and $w_0w_a$ model for the combination CMB+BAO+SNe with Union3 (black dashed line) as the ratio with respect to the best fit $\Lambda$CDM model for CMB+BAO. The gray band is the expected 1$\sigma$ errors with $\Delta \ell = 30$ for a measurement by an instrument similar to the Simons Observatory.}
\label{fig:ClEE}
\end{figure}

Any component that alters the expansion history and evolution of the perturbations at $z\approx 10^3$ risks destroying the good agreement between observations and theoretical models of the CMB anisotropy.  In particular, the E-mode of the CMB anisotropy polarization power spectrum ($C_\ell^{EE}$) is very sensitive to these effects.  Figure \ref{fig:ClEE} shows the ratio of $C_\ell^{EE}$ in our best-fitting, evolving DE cosmologies (Table \ref{tab:param_constraints}) to that the best-fitting $\Lambda$CDM cosmology. The grey bands indicate the forecasted $1\,\sigma$ errors from future measurements by the Simons Observatory \cite{SimonsObs,SimonsObsLAT}. Although the differences are very small, the fraction of EDE prefered by the current CMB+BAO data ($f_{\rm EDE} \approx 0.09$) could be confirmed or ruled out by this new CMB data set.

\begin{figure}
\centering
\resizebox{\columnwidth}{!}{\includegraphics[scale=1]{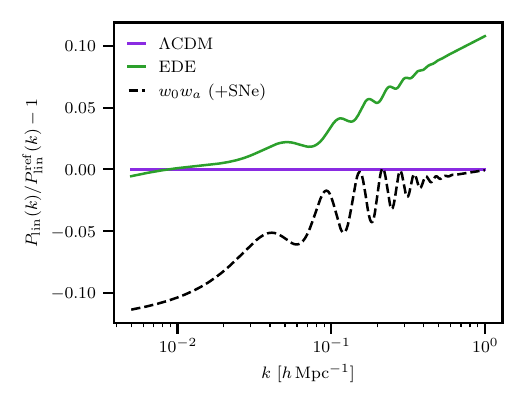}}
\caption{Ratio of the linear matter power spectrum at $z=1$ between our three model best fits to the combination of BAO+CMB data, plus SNe (Union3) for the $w_0w_a$ model.  We have chosen the $\Lambda$CDM model as the reference.}
\label{fig:Pk}
\end{figure}

\begin{figure}
\centering
\hspace{-1cm}\resizebox{1.1\columnwidth}{!}{\includegraphics{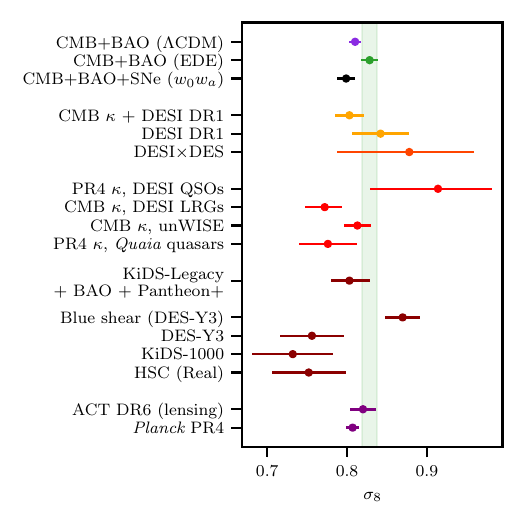}}
\caption{A selection of recent measurements of the low-$z$ amplitude of the power spectrum. The top points show the measurements from this paper, other measurements are taken from the literature (see text for citations). We caution the reader that most of these constraints are done outside the context of EDE models, and the redshifts and scales probed differ significantly so the points should be taken as indicative. For the cosmic shear datasets, we have converted from $S_8$ to $\sigma_8$ where necessary using their best fit $\Omega_m$, i.e.\ assuming motion along the degeneracy is largely unconstrained.}
\label{fig:sig8}
\end{figure}

Figure \ref{fig:Pk} compares the linear theory, matter power spectrum at $z=1$ for the best-fitting $\Lambda$CDM model, $w_0w_a$ model\footnote{Since we do not have a `microphysical' model for the $w_0w_a$ case, we compute the power spectra within the parameterized post-Friedman approximation, so the results assume the validity of that approximation.  However at $z\simeq 1$ the impact of dark energy is still small, so we believe this to be a reasonable approach.} and our best-fitting EDE model (the associated $\sigma_8$ values are given in Table \ref{tab:param_constraints}).  The fact that the best-fitting EDE model predicts a higher amplitude of clustering at late times than the best-fitting $\Lambda$CDM model has been noted (and explained) before, and has been used to disfavor the EDE model \cite{Smith:2019ihp,Hill2020,Ivanov20,Poulin23,McDonough24}.  While such a prediction increases the ``$S_8$-tension'' \cite{Abdalla22} we note that there have been some new results that bear upon this issue (Fig.~\ref{fig:sig8} shows a compilation of recent measurements \cite{Tristram24,Miyatake23,Heymans2021,DESY3,McCullough24,Wright25,Alonso23,Farren24,Sailer24,Belsunce25,Chen24,DESI2024.V.KP5,Maus25} compared to our theoretical predictions).  The 3D clustering amplitude inferred from modern perturbation theory models fit to the DESI data is larger than when these same class of models are fit to the older BOSS data \cite{DESI2024.V.KP5}.  New investigations into modeling assumptions within weak lensing surveys \cite{Arico23,Chen24,McCullough24,McCarthy24,Bigwood24,Ferreira24,Hadzhiyska25,Piccirilli25,Wright25} have also suggested that larger clustering amplitudes may be allowed by those data than previously thought.  Similarly, newer measurements of the CMB lensing auto-spectrum \cite{Qu24} are consistent with clustering amplitudes on the ``high end'' of the range.  For these reasons, we regard the prediction of EDE models for higher clustering to be concerning but provisionally allowed, pending a new analysis that self-consistently includes the impact of EDE.

Fortunately, new data will enable us to perform a precise test of this model in the very near future.  Wide-area cosmic shear catalogs are becoming available \cite{DECADE}, cosmology analysis of the final release of the Dark Energy Survey \cite{DES_Y6} is imminent, Euclid \cite{EUCLID18} is in orbit and expected to return data soon and the Legacy Survey of Space and Time on the Vera Rubin Observatory \cite{LSST09} will begin in short order.

Perhaps the strongest constraints will come from DESI DR2 data. Both the volume surveyed and the completeness of the DESI survey have improved dramatically compared to the DR1 data described above \cite{DESI2024.V.KP5}, enabling a 3D clustering analysis in which theoretical assumptions can be tightly controlled while still returning tight constraints.  Within the context of modern perturbation theory treatments, most cosmological constraining power comes from $k<0.1\,h\,\mathrm{Mpc}^{-1}$, with nuisance parameters degrading the impact of higher-$k$ terms.  The effective volume for DR2 is $2.2\times$ times larger than for DR1, suggesting differences such as those in Fig.~\ref{fig:Pk} should be distinguishable.  We intend to return to this question once the relevant data are unblinded within the DESI collaboration.

Finally we note that within the EDE model the changes to the power spectrum shape affect a wide range of redshifts.  Therefore high $z$ measures at smaller scales, from the Ly$\alpha$ forest \cite{Karacayli23,deBelsunce24} or future spectroscopic surveys \cite{WST,Spec-S5} for example, could potentially provide even stronger constraints on (or support for) the model.  We leave any such investigation to future work.

\section{Conclusions}
\label{sec:conclusions}

The recent DR BAO results from DESI \cite{DESI.DR2.BAO.cosmo} have increased the significance of the long-standing tension between acoustic waves measured in the CMB and BAO when interpreted within the $\Lambda$CDM model.  While still of modest statistical significance ($2.3\,\sigma$) it lends support to other lines of evidence arguing for a revision of the standard cosmological model (see \cite{DESI.DR2.BAO.cosmo} for further discussion).  Refs~\cite{DESI.DR2.BAO.cosmo,Y3.cpe-s1.Lodha.2025,Y3.cpe-s2.Elbers.2025} explored solutions to this puzzle that alter the behavior of DE at late times.  In this paper we have investigated solutions that instead alter the dynamics at high redshift.  As a specific example, we have studied the impact of early dark energy (EDE) in the form of a scalar field whose energy density contributes $\mathcal{O}(10\%)$ of the critical density at $z\sim 10^{3-4}$ before rapidly redshifting away.  Such a field briefly alters the expansion history, and thus the sound horizon scale that sets the `standard ruler' for BAO.

Our main focus has been on cosmological probes based on acoustic oscillations in the early and late Universe, specifically the CMB and BAO.  For the former we concentrate on Planck PR4 (including CMB lensing from Planck and ACT), while for the latter we use the newly released DESI DR2 BAO measurements.  While we discuss the inclusion of SNe data, it turns out that the acoustic data sets alone provide enough information to constrain the parameters of the EDE model (in contrast to the situation with the late-time model exemplified by $w_0w_a$).

We find that an EDE model can modestly ($\Delta\chi^2_{MAP}=7.4$ for 3 additional parameters) improve the agreement between the two key acoustic physics observables: the CMB and BAO.  Within the $\Lambda$CDM model the CMB data prefer larger distances to $z\simeq 0.3-1$ and a higher value of $\Omega_m$ than do the BAO data.  For dark energy fractions of $\approx 10\,\%$ the EDE model resolves these discrepancies, reducing the sound horizon at the drag epoch ($r_d$) by $3\,\%$ and increasing $H(z<10)$ by $\approx 5\%$.  This allows a model with $\Omega_m\simeq 0.3$ and $h\simeq 0.7$ to provide simultaneously good fits to the CMB and BAO data.  The final parameters are relatively close to those of the $\Lambda$CDM model fit to the same data combination, but in that case the two datasets are in (mild) tension and the agreement is more of a compromise on a model that neither set particularly prefers.

The set of EDE models that fit the combination of CMB and BAO data have higher $\omega_m$ and $H_0$ but lower $\Omega_m$ than $\Lambda$CDM.  The shift in $H_0$ is larger than in $\Omega_m$ such that $H(z)$ is increased by $\approx 5\%$ over the range $0<z<1$.  If distances are measured in $h^{-1}$Mpc, to incorporate this shift in $H_0$, then the distance scale is largely unchanged by the introduction of EDE.  By contrast the late-time solutions prefer higher $\Omega_m$ and lower $H_0$ (though $\omega_m$ is also modestly increased in this model).  The expansion history at low $z$ is altered much more significantly than in the EDE model.

Both the best-fitting EDE model and the best-fitting $w_0w_a$ model feature a dark energy component that first rises and then decays with cosmic time.  For both models the rise and decay take place over $\approx 1$ e-fold in expansion.  For EDE the DE component is never dominant, making up at most 10\% of the total energy density (Fig.~\ref{fig:fede_vs_z}).  For the $w_0w_a$ model that best fits the CMB+BAO+SNe data $H^{-1}\,d\ln\rho_{DE}/dt$ runs from $\approx 2$ at early time to $\approx -1$ today.  Naively extrapolating into the future, $\rho_{DE}>\rho_m$ for just under 1.5 e-folds of expansion.  Of course in the $w_0w_a$ model the action is near to the present day while for EDE it if confined to a few hundred thousand years after the big bang.  The EDE model additionally requires a cosmological constant component in order to match the accelerated expansion of the Universe today.

Assuming that we are not being misled by a statistical fluke or erroneous data, the emerging `acoustic tension' suggests a solution either in the early- or late-time Universe.  At present both are allowed by the data, but they make different predictions that allow them to be observationally distinguished.  For example, two of the SNe datasets prefer a late-time solution, with a rapid evolution in the dark energy below $z\simeq 0.5$.  Newer SNe data and improved BAO measurements from increased sky coverage will help to sharpen this distinction.  The values of $H_0$ in the EDE model are larger than in $\Lambda$CDM while those in $w_0w_a$ are smaller.  Improved distance-scale measurements will provide further discrimination.  The EDE models predict very modestly lower $\Omega_m$ than $\Lambda$CDM, while $w_0w_a$ prefers higher values.  At the level of the perturbations, both the CMB and future large-scale structure data could definitively settle this issue because the different models predict quite different CMB anisotropies and clustering.  In fact, the EDE model is already under tension from existing large-scale structure datasets, and DESI DR2 should provide significantly improved constraints in this regard.

While this work was under collaboration review within DESI, the Atacama Cosmology Telescope (ACT) collaboration released their latest results \cite{ACT_DR6_LCDM,ACT_DR6_ext}.  The mismatch in $\Omega_m$ between CMB and BAO, when interpreted within the $\Lambda$CDM framework, persists in the data \cite{ACT_DR6_LCDM} though it is slightly lessened in the combination of Planck and ACT.  The ACT data also allow $f_{\rm EDE}$ at the levels required to resolve the acoustic tension between CMB and BAO \cite{ACT_DR6_ext}.  The constraints tighten significantly if likelihoods preferring a lower clustering amplitude are included, but as we have argued above the strength of this tension is currently under debate within the community.

\section*{Data availability}
Data from the plots in this paper will be available on Zenodo\footnote{\url{xxxx}} as part of DESI’s Data Management Plan once the paper is accepted.

\section*{Acknowledgments}

This material is based upon work supported by the U.S. Department of Energy (DOE), Office of Science, Office of High-Energy Physics, under Contract No. DE–AC02–05CH11231, and by the National Energy Research Scientific Computing Center, a DOE Office of Science User Facility under the same contract. Additional support for DESI was provided by the U.S. National Science Foundation (NSF), Division of Astronomical Sciences under Contract No. AST-0950945 to the NSF’s National Optical-Infrared Astronomy Research Laboratory; the Science and Technology Facilities Council of the United Kingdom; the Gordon and Betty Moore Foundation; the Heising-Simons Foundation; the French Alternative Energies and Atomic Energy Commission (CEA); the National Council of Humanities, Science and Technology of Mexico (CONAHCYT); the Ministry of Science, Innovation and Universities of Spain (MICIU/AEI/10.13039/501100011033), and by the DESI Member Institutions: \url{https://www.desi.lbl.gov/collaborating-institutions}. Any opinions, findings, and conclusions or recommendations expressed in this material are those of the author(s) and do not necessarily reflect the views of the U. S. National Science Foundation, the U. S. Department of Energy, or any of the listed funding agencies.

The authors are honored to be permitted to conduct scientific research on I'oligam Du'ag (Kitt Peak), a mountain with particular significance to the Tohono O’odham Nation.

\bibliographystyle{biblio_style}
\bibliography{bibli}

\end{document}